\documentstyle[preprint,tighten,eqsecnum,aps,floats,psfig,epsfig,amsmath]{revtex}

\def\be{\begin{equation}}
\def\ee{\end{equation}}

\def\bea{\begin{eqnarray}}
\def\eea{\end{eqnarray}}

\def\e{\epsilon}
\def\s{\sigma}
\def\o{\omega}
\def\p{\varphi}

\def\Fa{\hat{a}}
\def\Fb{\hat{b}}
\def\FB{\hat{B}}

\def\Fd{\hat{d}}
\def\FD{\hat{D}}

\def\Ma{\check{a}}

\newcommand{\ev}[1]{\mbox{$\langle #1 \rangle$}}

\begin{document}

\title{Evolution of Entanglement Entropy in One-Dimensional Systems}
\author{Pasquale Calabrese$^1$ and John Cardy$^{1,2}$}
\address{$^1$Rudolf Peierls Centre for Theoretical Physics,
         1 Keble Road, Oxford OX1 3NP, U.K.\footnote{Address for
correspondence} \\
         $^2$All Souls College, Oxford.}
%
%
\maketitle
\begin{abstract}

We study the unitary time evolution of the entropy of entanglement of a 
one-dimensional system between the degrees of freedom in an interval of 
length  $\ell$ and its complement, starting from a pure state which is not
an eigenstate of the hamiltonian. 
We use path integral methods of quantum field theory as well as explicit 
computations for the transverse Ising spin chain. 
In both cases, there is a maximum speed $v$ of propagation of signals. 
In general the entanglement entropy increases linearly with time $t$ up 
to $t=\ell/2v$, after which it saturates at a value proportional to $\ell$,
the coefficient depending on the initial state.
This behavior may be understood as a consequence of causality.

\end{abstract}

\newpage

\section{Introduction}

Understanding the degree of entanglement of systems with many degrees 
of freedom, such as quantum spin chains, is currently one of the 
most challenging 
problems connecting quantum information science and statistical physics.
In the last few years several measures of entanglement have been 
proposed \cite{meas,Vidal}, and have been 
calculated (analytically in the simplest
instances, otherwise numerically) in the ground state of several
many-body systems. 
Despite a wide investigation, there is still no consensus 
on the correct measure of entanglement in the ground state of a many-body 
system. In this paper we will consider only the entanglement entropy,
since it is the measure most readily suited to analytic 
investigations, especially if quantum field theory methods 
are employed \cite{cc-04}.

The entanglement entropy is defined as follows.  
Suppose the whole system is in a pure quantum state $|\Psi\rangle$, 
with density matrix
$\rho=|\Psi\rangle\langle\Psi|$, and an observer A measures only a
subset $A$ of a complete set of commuting observables, while another
observer B may measure the remainder. A's reduced density matrix is
$\rho_A={\rm Tr}_B\,\rho$. The entanglement entropy is just the von
Neumann entropy $S_A=-{\rm Tr}_A\,\rho_A\log\rho_A$ associated with this
reduced density matrix. It is easy to check that $S_A=S_B$. For an
unentangled product state, $S_A=0$. Conversely, $S_A$ should be a
maximum for a maximally entangled state. 
One of the most striking features of the entanglement entropy is the 
universal behavior displayed at and close to a quantum phase transition.
In fact, close to a quantum critical point, where the correlation length 
$\xi$ is much larger than the lattice spacing $a$, 
the long-distance behavior of the correlations in the ground state of
a quantum spin chain are effectively described by a 1+1 dimensional 
quantum field theory.
At the critical point, where $\xi$ diverges, the field theory, if
Lorentz-invariant, is also  
a {\em conformal} field theory (CFT). 
In the latter case, it has been shown that, if $A$ is an interval of 
length $\ell$ in an infinite chain,  
$S_A\sim (c/3)\log\ell$ \cite{Holzhey,cc-04}, where $c$ is 
the central charge of the corresponding CFT.
In higher dimensions, the critical entanglement entropy $S_A$ scales like
the area separating $A$ and $B$ \cite{s-93,Vidal,cc-04}.
These results have been verified by analytic and numerical calculations on 
integrable quantum spin chains and in more complicated 
systems \cite{Vidal,jk-04,ijk-04,p-05,other,dur}. 
When the correlation length $\xi$ is large, but finite, it has been shown that
for $1$-dimensional systems $S_A= {\cal A}(c/6)\log\xi$ \cite{cc-04}, 
where ${\cal A}$ is the number of boundary points between $A$ and $B$
(the 1-d ``area''). This has been confirmed in some integrable models 
with ${\cal A}=1$ \cite{cc-04} and ${\cal A}=2$ \cite{ijk-04,p-05}.
Finally the dependence of the entanglement entropy on the central charge
has been the starting point for understanding the 
connection between irreversibility 
along renormalization group trajectories (Zamolodchikov's 
$c$-theorem\cite{Zam}) and loss of entanglement \cite{Vidal,loss}.

A less investigated situation is how (one measure of) entanglement evolves 
when the system is prepared in a state that is {\em not} an eigenstate of 
the corresponding hamiltonian.
Only a few works considered some entanglement measures in rather specific 
models \cite{outeq,dur}, thus universal features as in the ground state have 
not yet emerged. 
In this paper we mainly consider a scenario that is simply
realized experimentally, for example in ultracold atoms \cite{uc,sps-04}. 
Consider a quantum system at $T=0$ with an hamiltonian $H(\lambda)$ depending 
on a tunable experimental parameter $\lambda$ (for simplicity one may
think of the Ising chain in a transverse magnetic 
field $\lambda$ \cite{sach}, but the scenario is more general).
The system is prepared in a pure state $|\psi_0\rangle$,
which corresponds to an eigenstate of $H(\lambda_0)$ with 
$\lambda_0\neq \lambda$. Then, for example,
at time $t=0$ the parameter is suddenly quenched from 
$\lambda_0$ to $\lambda$. In general we expect that $|\psi_0\rangle$ 
is {\it not} an eigenstate of $H(\lambda)$, and thus the system evolves 
unitarily according to the equations of motion given by $H(\lambda)$, 
in the absence of any dissipation (as we assume throughout  the paper). 
In general, the system does not relax to its ground state,
as it is the case in the presence of dissipation.
Within this scenario we consider the time-evolution of 
entanglement entropy.

In this paper we concentrate on the case of one space dimension, using
two complementary methods which have been successfully applied to the
time-independent case. The first uses the path integral representation
and the methods of quantum field theory. It is straightforward formally to
write a path integral representation for the reduced density matrix
\begin{equation}
\rho_A(t)=
{\rm
Tr}_B\,e^{-iH(\lambda)t}|\psi_0\rangle\langle\psi_0|e^{iH(\lambda)t}
\end{equation}
and then, following \cite{cc-04}, to write the moments 
${\rm Tr}_A\,\rho_A^n(t)$ as path integrals over an $n$-sheeted surface with
branch cuts along A. To proceed further however, we assume that
$H(\lambda)$ corresponds to a continuum conformal field theory. As in
any continuum field theory, the energy fluctuations in the state
$|\psi_0\rangle$ are then divergent. The simplest way of regularizing
these while keeping the calculation tractable is to modify the state
according to the prescription
$|\psi_0\rangle\to e^{-\e H(\lambda)}|\psi_0\rangle$, which
filters out the high energy components. 

With this modification, we find that when $A$ is an interval of length
$\ell$, in the limit when $\ell/\e,t/\e\gg1$, the
entanglement entropy $S_A$ grows linearly with $t$, up to a time 
$t\sim\ell/2$ (in units where the speed of the
elementary excitations is unity). Thereafter 
it saturates at a value
$S_A\sim \pi c\ell/12\e$. These results, and the crossover time 
$t\sim\ell/2$, can be understood in terms of causality applied to
entangled left- and right-moving pairs of quasiparticles emitted
from the initial state. 


As a complement to the general conformal field theory calculations, we 
calculate the time-dependent entanglement entropy $S_A$ 
for the simplest integrable model: The Ising chain in a transverse magnetic 
field. The reason for this calculation is twofold. 
On the one hand, it is a practical check of the CFT results just summarized 
and can help understanding problems connected with the 
regularization $\e$. 
On the other, it allows exact calculations even in the non-critical 
(massive) phase, thus generalizing the CFT results.

We consider the time evolution from the ground state at a given transverse 
magnetic field $h_0$, according to the equation of motion given by 
$H(h)$ with $h\neq h_0$.
We obtain the remarkable analytic results (equation (\ref{Stinfinito})) that 
for $t\rightarrow\infty$, $S_A$ is proportional to $\ell$ for each pair 
of $h$ and $h_0$ (as long as $h\neq h_0$).
This confirms the CFT results and further shows that the linear 
dependence on $\ell$ for large times is not a peculiarity of the critical 
evolution. 
Thus for each given $h$, there is an arbitrary large entanglement entropy 
in the asymptotic state, contrarily to the ground-state case, where infinite 
entropy is allowed only at criticality.
Furthermore we obtain that $\lim_{\ell\rightarrow\infty} S_A/\ell$ 
is bounded from 
above by $2\log 2-1$ (the value found for $h_0=\infty,\,h=1$). 
This maximum value of the entanglement entropy is less than $\ell \log 2$,  
a rigorous bound for a generic state of $\ell$ spins one-half.
An interesting (and not understood) feature of the leading term of $S_A$ for 
$t\gg \ell\rightarrow\infty$ is that it is invariant under the exchange 
$h\leftrightarrow h_0$.

For finite times we find a linear increase of $S_A$ with $t$,
for all $h\neq h_0$, not only at criticality. 
The crossover between the linear and non-linear regime happens always at 
$t^*=\ell/2$, for large $\ell$.
However for $t>\ell/2$, $S_A$ does not immediately saturate to the 
asymptotic value, but is a slowly increasing function of the time.
In Sec. \ref{arg}, we ascribe this difference between lattice and CFT
results to the non-linear dispersion relation of quasiparticles emitted 
from the initial state. 
When both $h$ and $h_0$ get closer to the critical field,
the time dependence becomes sharper and more similar to the CFT results.

The layout of this paper is as follows. In the next section, we discuss the 
time dependence on the entanglement entropy within CFT, and in 
Sec. \ref{secIs} we consider the quantum Ising model in a transverse 
field (some cumbersome details of the calculation are postponed to 
two appendices).
In Sec. \ref{arg} we show that the results obtained may be explained in terms
of a physical picture invoking causality. 
Finally in Sec. \ref{disc} we discuss some points that 
need further investigation.

\section{Time evolution of entanglement entropy in quantum field theory}
\label{cftsec}

\subsection{Path integral formulation}

Suppose (in the notation of \cite{cc-04}) we prepare the system in
a state $|\psi_0(x)\rangle$ and unitarily evolve it with the
hamiltonian $H$. The matrix elements of the density matrix at time
$t$ are
\begin{equation}
\langle\psi''(x'')|\rho(t)|\psi'(x')\rangle=Z_1^{-1}
\langle\psi''(x'')|e^{-itH-\e H}|\psi_0(x)\rangle
\langle\psi_0(x)|e^{+itH-\e H}|\psi'(x')\rangle\,. \label{dm0}
\end{equation}
We have included damping factors $e^{-\e H}$
in such a way as to make the path
integral absolutely convergent. We shall see at the end of the
calculation whether it is justified to remove them.

\begin{figure}[ht]
\centering
\includegraphics[width=8cm]{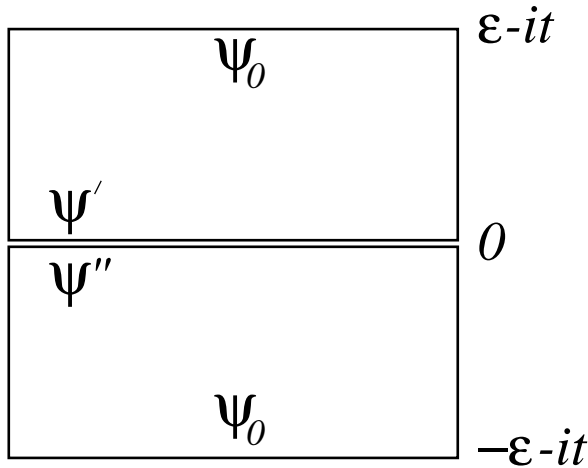}
\caption{Space-imaginary time regions for the density matrix in (\ref{dm0})}
\label{figpi}
\end{figure}

Each of the factors may be represented by an analytically
continued path integral in imaginary time: the first one over
fields $\psi(x,\tau)$ which take the boundary values $\psi_0(x)$
on $\tau=-\e-it$  and $\psi''(x)$ on $\tau=0$, and the second with
$\psi(z,\tau)$ taking the values $\psi'(x)$ on $\tau=0$ and
$\psi_0(x)$ on $\tau=\e-it$. This is illustrated in
Fig.~\ref{figpi}.

The normalization factor
\begin{equation}
Z_1=\langle\psi_0(x)|e^{-2\e H}|\psi_0(x)\rangle\,,
\end{equation}
which guarantees that Tr\,$\rho=1$ is given by the euclidean path
integral over imaginary time $2\e$, with initial and final
conditions both equal to $\psi_0(x)$. This is the same as sewing
together the two edges in Fig.~\ref{figpi} along $\tau=0$.

As in \cite{cc-04}, the reduced density matrix $\rho_A(t)$ is giving
by sewing together along $\tau=0$ only those parts of the $x$-axis
corresponding to points in B, leaving open slits along A, and
Tr$\,\rho_A^n$ is given by sewing together $n$ copies of this in a
cyclic fashion. Thus
\begin{equation}
{\rm Tr}\,\rho_A^n=Z_n/Z_1^n\,,
\end{equation}
where $Z_n$ is the path integral on an $n$-sheeted surface, where
the edges of each sheet correspond to imaginary times $-\tau_1$
and $\tau_2$, analytically continued to $\tau_1=\e+it$ and
$\tau_2=\e-it$, and the branch points lie along $\tau=0$ at
the boundaries points between A and B.
Finally, as derived in \cite{cc-04}, the entanglement entropy is given by
\be
S_A=\left.-\frac{\partial}{\partial n} {\rm Tr}\,\rho_A^n\right|_{n=1}\,.
\ee

\subsection{Conformal mapping}

Now consider the case when $H$ is critical and the field theory is
a CFT. Under the renormalization group any translationally
invariant boundary condition is supposed to flow into a boundary
fixed point, satisfying conformal boundary conditions. Thus we may
assume that the state $|\psi_0\rangle$ corresponds to such
boundary conditions on sufficiently long length scales. However
this means that the conclusions will be valid only asymptotically.

For real $\tau$ the strip geometry described above may be obtained
from the upper half-plane by the conformal mapping
$w=(2\e/\pi)\log z$, with the images of the slits lying along
${\rm arg} z=\pi\tau_1/2\e$. The result for $Z_n/Z_1^n$ in the
upper half $z$-plane, for an arbitrary number of branch points,
was given in \cite{cc-04}. It has the form of a correlation
function of local operators $\Phi_{\pm n}$ located at the branch
points, which transforms simply under conformal mappings.

First consider the case when A is a slit of length $\ell$ (and B
is the rest of the real axis.) In that case we can take the image
of the branch points to be $z_1=\rho^{-1}e^{i\pi\tau_1/2\e}$
and $z_2=\rho e^{i\pi\tau_1/2\e}$ where
$\rho=e^{\pi\ell/4\e}$. From \cite{cc-04} in the $z$-plane we
have
\begin{equation}\label{zzzz}
{\rm Tr}\,\rho_A^n=\langle\Phi_n\Phi_{-n}\rangle\sim
c_n\left(\frac{|z_1-\bar z_2||z_2-\bar z_1|}{|z_1-z_2||\bar
z_1-\bar z_2||z_1-\bar z_1||z_2-\bar z_2|}\right)^{2n\Delta_n}\,,
\end{equation}
where $\Delta_n=(c/24)(1-1/n^2)$ in the notation of \cite{cc-04}.
This formula is supposed to be valid in the euclidean region, when
all the separations are large. Under the conformal mapping,
\begin{equation}
\langle\Phi_n(w_1)\Phi_{-n}(w_2)\rangle=
|w'(z_1)w'(z_2)|^{-2n\Delta_n}\langle\Phi_n(z_1)\Phi_{-n}(z_2)\rangle\,.
\end{equation}
After some algebra and continuing to $\tau_1=\e+it$, we find
\begin{equation}
{\rm Tr}\,\rho_A^n(t)\sim c_n(\pi/2\e)^{4n\Delta_n}
\left(\frac{e^{\pi\ell/2\e}+e^{-\pi\ell/2\e}+2\cosh(\pi t/\e)}
{(e^{\pi\ell/4\e}-e^{-\pi\ell/4\e})^2 \cosh^2(\pi
t/2\e)}\right)^{2n\Delta_n}\,.
\end{equation}
In the case where $\ell/\epsilon$ and $t/\epsilon$ are large this
simplifies to
\begin{equation}
c_n(\pi/2\e)^{4n\Delta_n}\left(\frac{e^{\pi\ell/2\e} +e^{\pi
t/\e}} {e^{\pi\ell/2\e}\cdot e^{\pi t/\e}}\right)^{2n\Delta_n}\,.
\end{equation}

Differentiating wrt $n$ to get the entropy,
\be
S_A(t)\sim \left\{
\begin{matrix} 
\displaystyle      \frac{\pi ct}{6\e}    \qquad(t<\ell/2)\;,\cr\cr
\displaystyle      \frac{\pi c\,\ell}{12\e}\qquad(t>\ell/2)\,,
\end{matrix}\right.
\label{SAt2}  
\ee
that is $S_A(t)$ increases linearly until it saturates at
$t=\ell/2$. The sharp cusp in this asymptotic result is rounded
over a region $|t-\ell/2|\sim\e$.

However we see that $\e$ enters in an essential way. The reason is
that, in a continuum field theory (as compared with a quantum spin
model) a state like $|\psi_0\rangle$ has infinitely large mean
energy (as well as divergent energy fluctuations). In order to
make sense of the result it is necessary to filter out the
high-energy components of the state. Within the path integral
approach, this is most easily enforced with a cut-off function
$e^{-\e E}$. To compare with results from a lattice spin model we
should presumably take $\e$ to be of the order of the lattice
spacing. The linear behavior in $t$ for $t<\ell/2$, the break at
$t=\ell/2$, and the saturation at a value $\propto\ell$ all agree
with our exact results for the transverse Ising spin chain, in
Sec.~\ref{secIs}, although there are other differences in detail.

\section{Dynamics of entanglement entropy in the quantum Ising chain}
\label{secIs}

As a complement to the general CFT calculation just presented, in this
section we describe how analogous results can be found in an analytically 
tractable model. We consider the Ising spin chain in a transverse 
magnetic field, which has a quantum phase transition between a ferromagnetic
and a (quantum) paramagnetic phase.

The model is defined by the hamiltonian
\be
H_I(h)=-\frac{1}{2}\sum_j [\s^x_j\s^x_{j+1}+ h \s^z_j]\,,
\label{HI}
\ee
where $\sigma^{x,z}_j$ are the Pauli matrix acting on the spin at the site 
$j$ of an infinite chain. The quantum critical point is located 
at $h=1$ \cite{sach}. 
We consider the time evolution from an initial state $|\psi_0\rangle$ 
that is an eigenstate of $H_I$ for a field $h_0\neq h$. 
This experimentally means quenching at $t=0$ the magnetic field from $h_0$ 
to $h$. We consider only the case $h,h_0\geq 1$. 
The generalizations to the case $h,h_0<1$ and to more general spin chains, 
such as the XY model \cite{sach} are straightforward and we will not 
consider them here.

The determination of the time-dependent state 
$|\psi(t)\rangle=e^{-i H_I(h) t}|\psi_0\rangle$ (and consequently of the 
entanglement entropy) proceeds with the Jordan-Wigner transformation in terms
of Dirac or Majorana fermionic operators. 
All the details of these computations can be found in the Appendix \ref{app}.

The final result is that the time-dependent entanglement entropy for 
$\ell$ 
consecutive spins in the chain can be obtained (analogously to the ground
state case \cite{Vidal}) from the correlation matrix of the Majorana 
operators
\be
\Ma_{2l-1} \equiv \left( \prod_{m<l} \sigma_m^z \right) \sigma_l^x, \qquad
\Ma_{2l} \equiv \left( \prod_{m<l} \sigma_m^z \right) \sigma_l^y.
\label{eq:aa}
\ee
We introduce the matrix $\Gamma^A_\ell$ through the relation
 $\ev{\Ma_m\Ma_n} = \delta_{mn} + i {\Gamma^A_\ell}_{mn}$ with $1\leq m,n\leq \ell$.
It has the form of a block Toeplitz matrix
\be
\Gamma^A_\ell = \left[
 \begin{array}{ccccc}
\Pi_0  & \Pi_{-1}   &   \cdots & \Pi_{1-\ell}  \\
\Pi_1 & \Pi_0   & &\vdots\\

\vdots&  & \ddots&\vdots  \\
\Pi_{\ell-1}& \cdots  & \cdots  & \Pi_0 
\end{array}
\right], ~~~ \Pi_l = \left[\begin{array}{cc}
-f_l    & g_l \\
-g_{-l} & f_l
\end{array}
\right]\,.
\label{eq:GammaAL}
\ee
with
\bea
g_l &=&  \frac{1}{2\pi} \int_0^{2\pi} d\p e^{-i\p l} e^{-i\theta_\p}
(\cos \Phi_\p-i\sin \Phi_\p \cos 2\e_\p t)
\,,\nonumber\\
f_l&=& \frac{i}{2\pi} \int_0^{2\pi} d\p e^{-i\p l} \sin \Phi_\p \sin 2 \e_\p t
\,,
\label{eq:g2}
\eea
where
\bea
\e_\p &=& \sqrt{(h-\cos{\p})^2 +  \sin^2{\p}}\,,\nonumber\\
\e_\p^0 &=& \sqrt{(h_0-\cos{\p})^2 +  \sin^2{\p}}\,,\nonumber\\
e^{-i\theta_\p}&=& \frac{\cos\p-h-i\sin\p}{\e_\p}\,,\nonumber\\
\sin \Phi_\p&=&\frac{\sin\p(h_0-h)}{\e_\p \e_\p^0}\,, \nonumber\\
\cos \Phi_\p&=&\frac{1-\cos\p(h+h_0)+h h_0}{\e_\p \e_\p^0}\,.
\eea
Calling the eigenvalues of $\Gamma^A_\ell$ as $\pm i \nu_m$, $m=1\dots \ell$,
the entanglement entropy is $S=\sum_{m=1}^\ell H(\nu_m)$ where $H(x)$ is
\be
H(x)= -\frac{1+x}{2} \log\frac{1+x}{2}  -\frac{1-x}{2} \log\frac{1-x}{2}\,.
\label{Hx}
\ee
The derivation of all these results is in the Appendix \ref{app}.

Note that the first trivial result is that for $t=0$ and for $h=h_0$,
the correlation matrix reduces to that of the ground state with field $h_0$,
as it should.

\subsection{Analytical results for $t\rightarrow\infty$}

In the limit $t\rightarrow\infty$ the matrix $\Gamma_\ell^A$ simplifies 
in such a way that analytical calculations are feasible for all $h,h_0$.
In fact, for $t\rightarrow \infty$, the time-dependent sine and cosine 
appearing in Eqs. (\ref{eq:g2}) average to zero, and $\Gamma_\ell^A$ has 
a simpler form with $f_l=0$ and 
\be
g_l=\frac{1}{2\pi} \int_0^{2\pi} d\p e^{-i\p l} 
\frac{e^{-i \p}-h}{\e_\p^2 \e_\p^0}(1-\cos\p(h+h_0)+h h_0)\,.
\ee
Thus it can be written as a block-Toeplitz matrix with
\be 
\Pi_l=\frac{1}{2\pi}\int_0^{2\pi} d\p e^{-i\varphi l} \Pi(\p)\,,
\ee
and
\be
\Pi(\p)=
 \left[\begin{array}{cc}
0 & \displaystyle{\frac{e^{-i\p}-h}{\e_\p^2}\,
\frac{1-\cos\p(h+h_0)+h h_0}{e^0_\p}} \\
- \displaystyle{\frac{e^{i\p}-h}{\e_\p^2}\,
\frac{1-\cos\p(h+h_0)+h h_0}{e^0_\p}} & 0
\end{array}
\right]\,.
\ee
In this case, following the method used for the ground state in 
Refs. \cite{jk-04,ijk-04} it is possible to calculate the leading behavior
of $S_\ell(\infty)$ for large $\ell$. 
Let us introduce \cite{jk-04,ijk-04}
\bea
\tilde{\Gamma}_\ell&=& i \lambda I_\ell -\Gamma^A_\ell\,,\\
\tilde\Pi(\p)&=&i\lambda I_1-\Pi(\p)\,,\\
e(y,x)&=&-\frac{y+x}{2} \log\frac{y+x}{2}-\frac{y-x}{2} \log\frac{y-x}{2}\,,
\eea
where $I_\ell$ is the $2\ell\times2\ell$ identity matrix.

The determinant of $\tilde{\Gamma}_\ell$ is
\be
D_\ell(\lambda)\equiv\det\tilde{\Gamma}_\ell=(-1)^\ell \prod_j^\ell (\lambda^2-\nu_j^2)\,,
\ee
so we can use Cauchy theorem to rewrite the entropy as \cite{ijk-04}
\be
S_\ell=\lim_{\e\rightarrow0^+}\frac{1}{4\pi i}
\oint_{C}d\lambda \,e(1+\e,\lambda)\frac{d}{d\lambda}\ln D_\ell(\lambda)\,,
\label{Kformula}
\ee
where the contour $C$ depends on the parameter $\e$ and includes the 
interval $[-1,1]$, so to encircle all the zeros of $D_\ell(\lambda)$.
This guarantees that the branch points of $e(1+\e,\lambda)$ are 
outside $C$ and $e(1+\e,\lambda)$ is analytic there.
As $\e$ goes to zero the contour approaches the interval $[-1,1]$.

\begin{figure}[tb]
\centerline{\epsfig{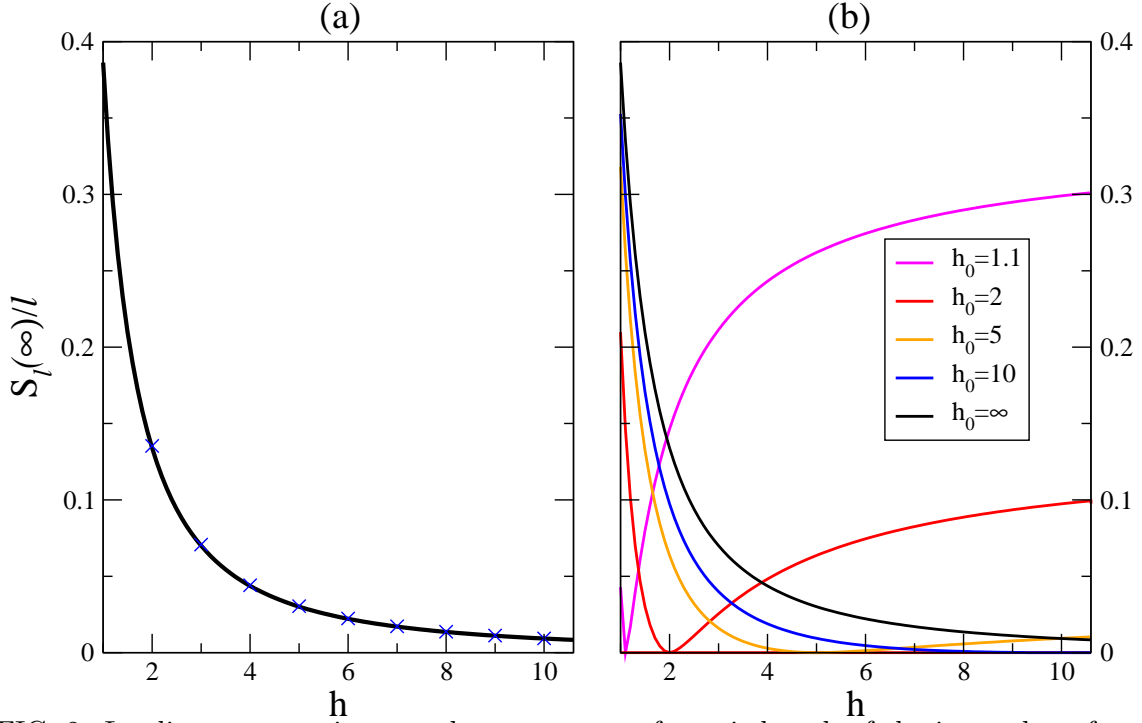}}
\caption{Leading asymptotic entanglement entropy for unit length of the 
interval, as function of $h$ Eq. (\ref{Stinfinito}).
(a) The case $h_0=\infty$ (full line) is compared with the numerical 
results (crosses) obtained by diagonalizing $\Gamma^A_\ell$ for $\ell=200$ 
and using the exact coefficients Eq. (\ref{glh0infinity}). 
(b) As in (a), but for several finite values of $h_0$. }
\label{Stinf}
\end{figure}

An old short overview about theorems concerning this type of
matrices can be found in \cite{am-74}. 
In particular, it is known that  the large $\ell$ asymptotic 
behavior of the determinant of a block Toeplitz matrix $T_\ell$ 
(under some hypothesis satisfied in our case $t=\infty$) is 
given by the generalization of the Szego lemma \cite{am-74}
\be
\log\det T_\ell=\ell \frac{1}{2\pi}\int_0^{2\pi} d\p\log[\det a(\p)]+O(\log \ell)\,,
\ee
where $a(\p)$ is the generating matrix-function of $T_\ell$ (in our case 
$\tilde\Pi(\p)$).
Thus 
\be
\log\det \tilde{\Gamma}_\ell=
\frac{\ell}{2\pi}\int_0^{2\pi}d\p \log[\det \tilde{\Pi}(\p)]+O(\log \ell)\,,
\ee
so (we omit $O(\log \ell)$)
\be
\frac{d\log\det \tilde{\Gamma}_\ell (\lambda)}{d\lambda}=
\frac{\ell}{2\pi}\int_0^{2\pi}d\p 
\frac{2\lambda}{\lambda^2-\frac{(1-\cos\p (h+h_0)+h h_0)^2}{(h^2+1-2h\cos\p)(h_0^2+1-2h_0\cos\p)}}\,.
\ee
Inserting this expression in Eq. (\ref{Kformula}) we get
\bea
S_\ell(\infty)&=&\lim_{\e\rightarrow0^+}\frac{1}{4\pi i}
\oint_{C}d\lambda \,e(1+\e,\lambda)
\frac{d}{d\lambda}\ln\det \tilde{\Gamma}_\ell(\lambda) =\nonumber\\
&=&\lim_{\e\rightarrow0^+}
\frac{1}{4\pi i} \oint_{C}d\lambda \,e(1+\e,\lambda)
\frac{\ell}{2\pi}\int_0^{2\pi}d\p 
\frac{2\lambda}{\lambda^2-\frac{(1-\cos\p (h+h_0)+h h_0)^2}{(h^2+1-2h\cos\p)(h_0^2+1-2h_0\cos\p)}}\,.
\eea
Changing the order of the integrals, using the residue theorem, and the
property $e(y,x)=e(y,-x)$ we get\footnote{In a similar way one can also 
calculate the the moments ${\rm Tr}\rho_A^n$.
In fact, using the results reported in the Appendix, it is straightforward
to show that $\log {\rm Tr}\rho_A^n$ is given by equation
(\ref{Stinfinito}) with $H$ replaced by $H_n=\log\{[(1-x)/2]^n+[(1+x)/2]^n\}$.
It follows that ${\rm Tr}\rho_A^n\propto e^{-f_n(h,h_0)\ell}$, 
where $f_n$ satisfies $f_n(h,h_0)=f_n(h_0,h)$.}
\bea
S_\ell(\infty)&=&\lim_{\e\rightarrow0^+}
\frac{\ell}{2\pi}\frac{1}{4\pi i} \int_0^{2\pi}d\p 
\oint_{C}d\lambda \,e(1+\e,\lambda) 
\frac{2\lambda}{\lambda^2-\frac{(1-\cos\p (h+h_0)+h h_0)^2}{(h^2+1-2h\cos\p)(h_0^2+1-2h_0\cos\p)}}\nonumber\\
&=&\frac{\ell}{2\pi}\int_0^{2\pi} d\p H\left(
\frac{1-\cos\p (h+h_0)+h h_0}{\sqrt{(h^2+1-2h\cos\p)(h_0^2+1-2h_0\cos\p)}}
\right)\,.
\label{Stinfinito}
\eea

Let us discuss this result. 
First, independently of $h$ and $h_0$,  $S_\ell(\infty)/\ell$ is 
non-vanishing (as long as $h\neq h_0$). 
Thus the asymptotic entanglement entropy for the Ising chain is always linear
in $\ell$, not only at the critical point as we already knew from CFT.  
For $h=1$, $h_0=\infty$ it reduces to $S_\ell(\infty)/\ell=2 \log2-1$, 
as it can be understood from a very simple argument reported in 
Appendix \ref{appb}. 
This is the maximum value that $S_\ell(\infty)/\ell$ can asymptotically reach.
In Fig. \ref{Stinf} (a) we depict $S_\ell(\infty)/\ell$ for a quench from 
$h_0=\infty$ as a function of $h$: It is a monotonous decreasing function of 
$h$. 
In Fig. \ref{Stinf} (b) we depict $S_\ell(\infty)/\ell$ as function of $h$, for 
several $h_0$. 
For $h<h_0$ it is a decreasing function, while for $h>h_0$ it increases, and 
obviously it is zero at $h=h_0$.
A further curious (and not understood) feature is that $S_\ell(\infty)$ is 
symmetric under the exchange of $h$ and $h_0$.

Note that in Eq. (\ref{Stinfinito}) corrections of the order $O(\log \ell)$
are neglected. These corrections are smaller than the leading term, but can be
arbitrarily large in the limit $\ell\rightarrow\infty$. 
They are in principle obtainable using generalizations 
of the Fisher-Hartwig conjecture \cite{ijk-04}, and are not expected 
to be symmetric under $h\leftrightarrow h_0$.

\subsection{Some numerical results}

Now, we present the time-dependence of the entanglement entropy, as 
obtained by numerical diagonalization of $\Gamma^A_\ell$.
We consider several quenches in the next subsections.

\begin{figure}[t]
\centerline{\epsfig{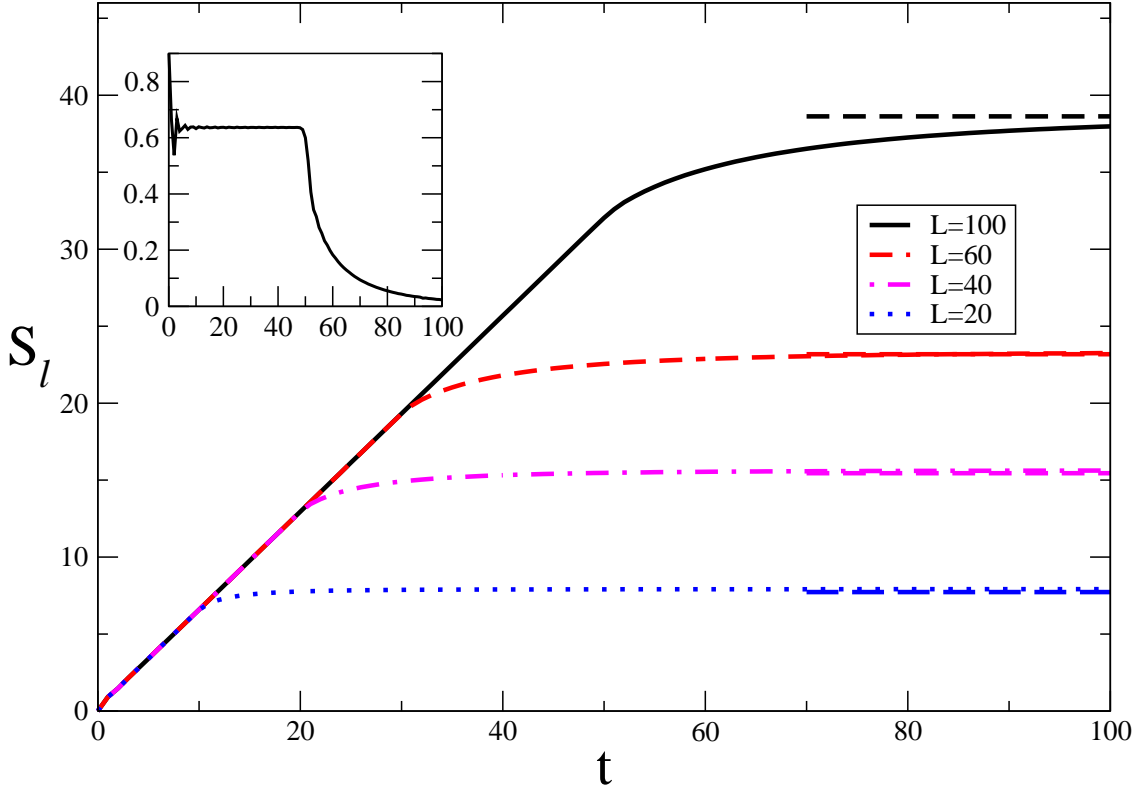}}
\caption{Entanglement entropy for the quench from $h_0=\infty$ to $h=1$, for 
various $\ell$. 
The dashed lines are the leading asymptotic results for large $\ell$, cf. 
Eq. (\ref{Stinfinito}).
The inset shows the derivative with respect to the time of $S_{100}(t)$.}
\label{SLL}
\end{figure}

\subsubsection{Quench from $h_0=\infty$ to $h=1$.}

We start our numerical analysis from the case in which the system is 
quenched from a completely uncorrelated state ($h_0=\infty$) to the critical 
point ($h=1$). 
The time evolution is governed by the critical hamiltonian, so
the results we obtain are expected to be well-described by CFT.

The numerical results, as obtained by diagonalizing numerically 
$\Gamma^A_\ell$, are shown in Fig. \ref{SLL} for $\ell=20,40,60,100$. 
For all $\ell$ considered, $S_\ell(t)$ exhibits just two time-regimes: 
a linear one for $t<t^*= \ell/2+O(\ell^0)$, and a nonlinear one for $t>t^*$.
The crossover between the two regimes happens 
at $t^*\simeq \ell/2$, in agreement with the CFT prediction. 
Note (inset of Fig. \ref{SLL}) that the derivative with respect to the 
time of $S_{100}(t)$ is practically constant for $t<50$, apart from an 
expected short-time non-universal behavior (that should disappear in the 
limit $\ell\rightarrow\infty$).
For $t<t^*$, all the $S_\ell(t)$ fall on the same master curve, 
independently of $\ell$.
However, the entropy does not saturate to the asymptotic value
exactly at $t^*$, as in CFT. 
In fact for large $t$, $S_\ell(t)$ is a slowly increasing function, even in 
the limit of large $\ell$. 

The dashed lines in Fig. \ref{SLL} are the leading asymptotic results for 
large $\ell$ as given by Eq. (\ref{Stinfinito}).
The actual asymptotic value (obtained by numerical diagonalization) is
always slightly larger than (\ref{Stinfinito}), showing that the first 
correction (most probably of the order $O(\log \ell)$) is positive.

\subsubsection{Quench from $h_0>1$ to $h=1$.}

\begin{figure}[t]
\centerline{\epsfig{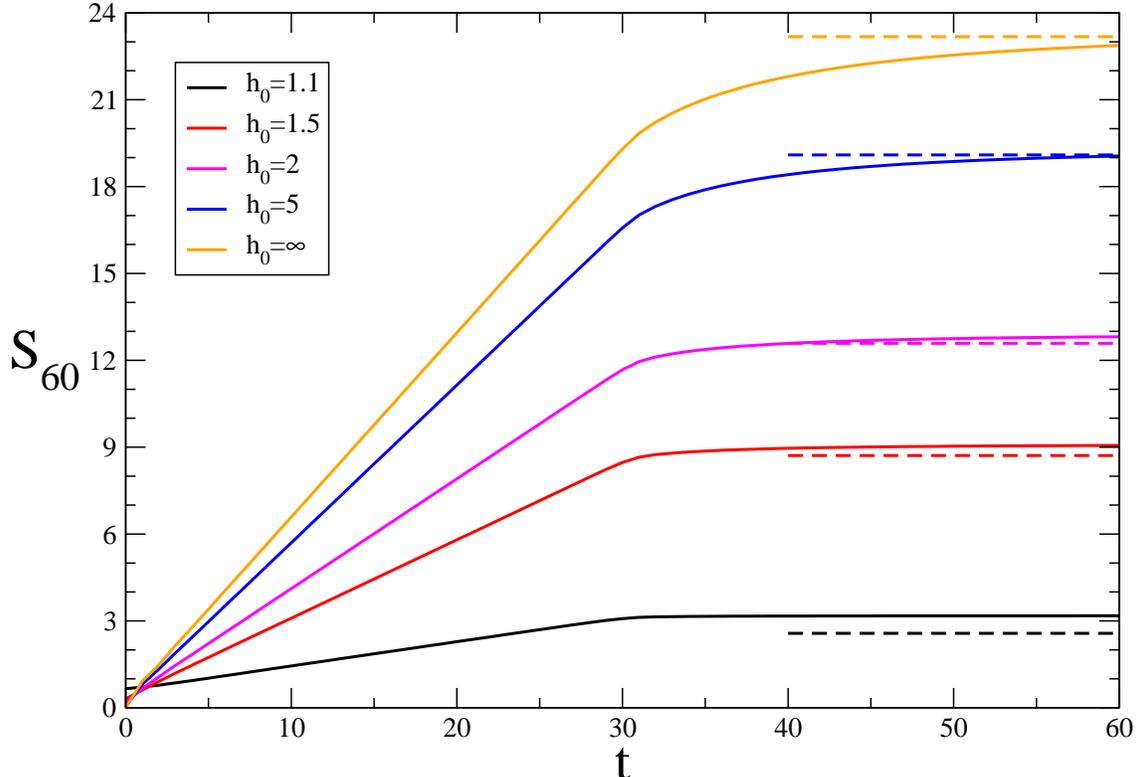}}
\caption{$S_{60}(t)$ for the quench from 
$h_0=\infty,\,5,\,2,\,1.5,\,1.1$ to $h=1$.
The dashed lines are the leading asymptotic results for large $\ell$ 
Eq. (\ref{Stinfinito}).
}
\label{SLh0}
\end{figure}

To understand if the disagreement between CFT and the exact 
result for $t>t^*$ can be attributed simply to the particular initial 
condition ($h_0=\infty$) or to deeper reasons, we consider now the case of 
a generic initial condition (i.e. finite $h_0$), always evolving according to 
the critical hamiltonian.

The numerical results for various $h_0$ (at fixed $\ell=60$) are displayed 
in Fig. \ref{SLh0}.
Even in this case the curves show for $t<t^*$ a linear regime, in 
agreement with CFT.
However, decreasing $h_0$ the time dependence for $t>t^*$ becomes sharper, 
i.e. $S_\ell(t)$ reaches the asymptotic value much faster than 
in the case $h_0=\infty$.
Thus the curves look more similar to the CFT result.

In Fig. \ref{SLh0} the dashed lines are the leading asymptotic results for 
large $\ell$ as given by Eq. (\ref{Stinfinito}).
Note that decreasing $h_0$, the positivity of the corrections to 
Eq. (\ref{Stinfinito}) is more and more apparent. 
In this case, it can be partially attributed to a non-vanishing value for 
$t=0$, that results in a {\em finite} shift of the entire curve.

\subsubsection{Quench from $h_0=\infty$ to $h>1$.}

\begin{figure}[t]
\centerline{\epsfig{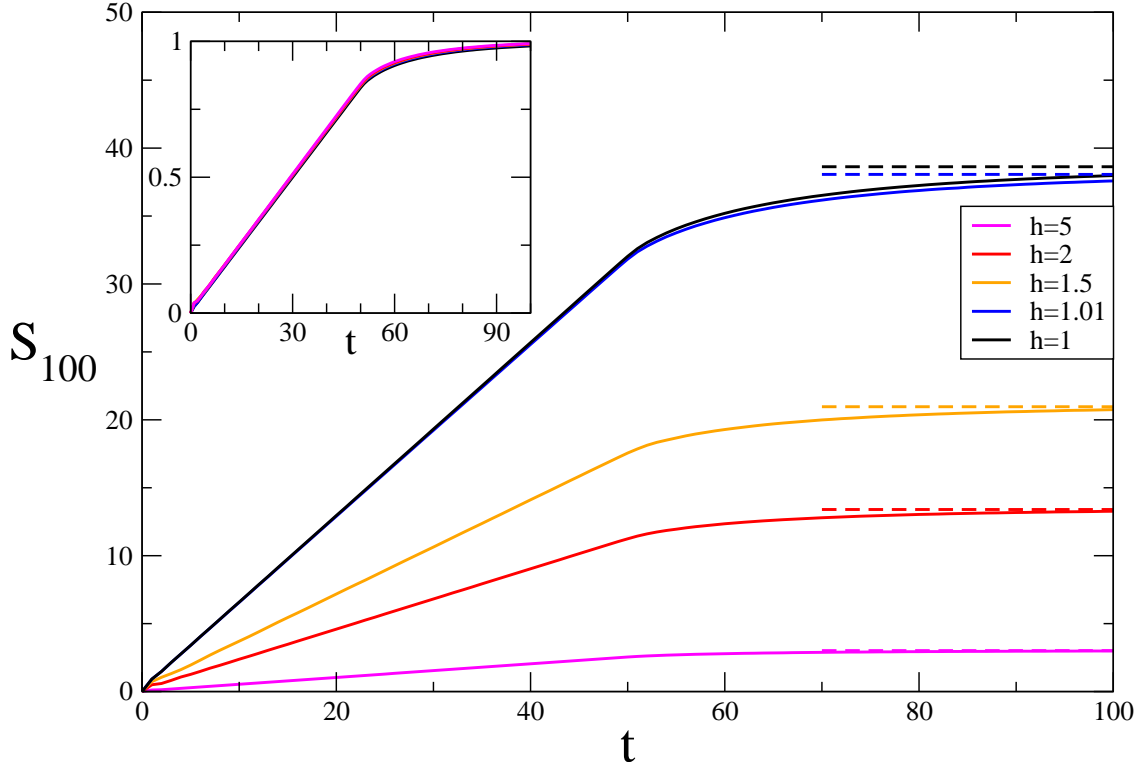}}
\caption{$S_{100}(t)$ for the quench from $h_0=\infty$ to $h=5,2,1.5,1.01,1$.
The dashed lines are the leading asymptotic results for large $\ell$ 
Eq. (\ref{Stinfinito}).
The inset shows the rescaling of the curves, according to the asymptotic 
value $S_{100}(\infty)$.}
\label{SLh}
\end{figure}

In the lattice calculation we are performing we are not restricted to
study the critical evolution. Thus, to understand how the entanglement 
entropy is affected by a non-critical evolution, we can simply diagonalize
$\Gamma_A^\ell$ for $h\neq1$.
As a first case, we consider a quench from $h_0=\infty$ to a 
non critical value $h>1$.
The numerical results, for fixed $\ell=100$, are shown in Fig. \ref{SLh}. 
The curves appear very similar to each other, and in particular to the 
critical case. There are two time regimes as in the critical case.
The crossover from the linear to non-linear regime is always at 
$t^*\simeq \ell/2$, independently of $\ell$ and $h$.  
A quite unexpected fact is that the all the curves fall 
on the same master curve, when rescaled according to $S_{100}(\infty)$
(see the inset in Fig. \ref{SLh}). 
We have no simple explanation for this.

\subsubsection{Quench from $h_0>1$ to $h>1$.}

\begin{figure}[t]
\centerline{\epsfig{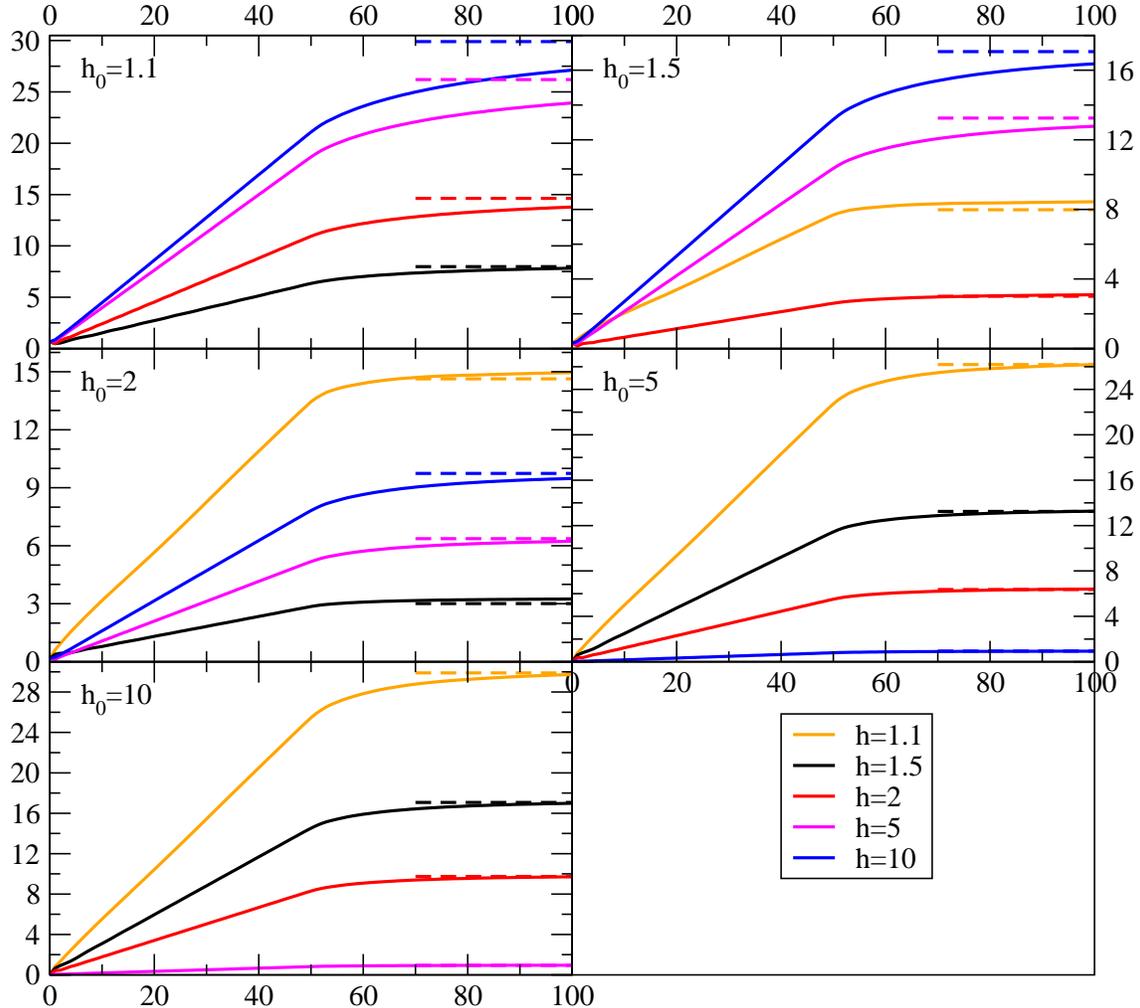}}
\caption{$S_{100}(t)$ for a general quench from some generic $h_0$ to several 
$h$.
The dashed lines are the leading asymptotic results for large $\ell$ 
Eq. (\ref{Stinfinito}).}
\label{SLhho}
\end{figure}

Finally we consider the most general quench into the paramagnetic phase.
Some numerical results for $\ell=100$ are shown in Fig. \ref{SLhho}. 
Due to the double variation of the parameters, these figures are not easily 
readable. The most interesting features are:
\begin{itemize}
\item The crossover from the linear to non-linear regime is 
always at $t^*\simeq \ell/2$.

\item At fixed $h_0$, when the curves are rescaled according to 
$S_\ell(t=\infty)$, they do {\em not} fall on the same curve as for $h_0=\infty$.

\item The dashed lines in Fig. \ref{SLhho} are the leading asymptotic results
for large $\ell$ as given by Eq. (\ref{Stinfinito}).
From the figures it is not always clear whether Eq. (\ref{Stinfinito}) is 
always smaller than the actual value (especially for small $h_0$).
We checked in all the cases not enough clear graphically (i.e. $h_0=1.1$ with
$h=10,5,2$ and $h_0=1.5$ with $h=10,5$) that for longer non-depicted times
the curves finally cross the value given by Eq. (\ref{Stinfinito}).
This shows quite firmly that for all $h,h_0$ the leading corrections to 
Eq. (\ref{Stinfinito}) are positive.
\item The actual asymptotic values for $t\rightarrow\infty$ for the cases 
with $h$ and $h_0$ interchanged are close, but not exactly equal as in 
Eq. (\ref{Stinfinito}). 
We checked that, for larger $\ell$, their difference is almost constant and 
so can be imputed to finite (or $O(\log \ell)$) corrections that consequently 
have not the symmetry under the exchange of $h$ and $h_0$ as the leading term.
\end{itemize}

\section{Physical Interpretation}
\label{arg}

The qualitative, and many of the quantitative, features of $S_A(t)$
found in the last two sections may be understood physically as follows.
We emphasize that this is not an ab initio calculation but rather a
simplified model which allows us to explain these observations.
The initial state $|\psi_0\rangle$ has a very high energy relative to
the ground state of the hamiltonian $H(\lambda)$ which governs the 
subsequent time
evolution, and therefore acts as a source of quasiparticle excitations.
Particles emitted from different points (further apart than the
correlation length in the initial state) are incoherent, but pairs of
particles moving to the left or right from a given point are highly
entangled. We suppose that the cross-section for producing such a pair
of particles of momenta $(p',p'')$ is $f(p',p'')$, and that, once they
separate, they move classically. 
This will of course depends on $H(\lambda)$ and the state $|\psi_0\rangle$,
and in principle is calculable, but we made no strong assumptions on its
form. If the quasiparticle dispersion relation
is $E=E(p)$, the classical velocity is $v(p)=dE/dp$. We assume that
there is a maximum allowed speed which is taken to be 1, that is
$|v(p)|\leq 1$. A quasiparticle of momentum $p$ produced at $x$ is
therefore at $x+v(p)t$ at time $t$, ignoring scattering effects.

\begin{figure}[t]
\centering
\includegraphics[width=10cm]{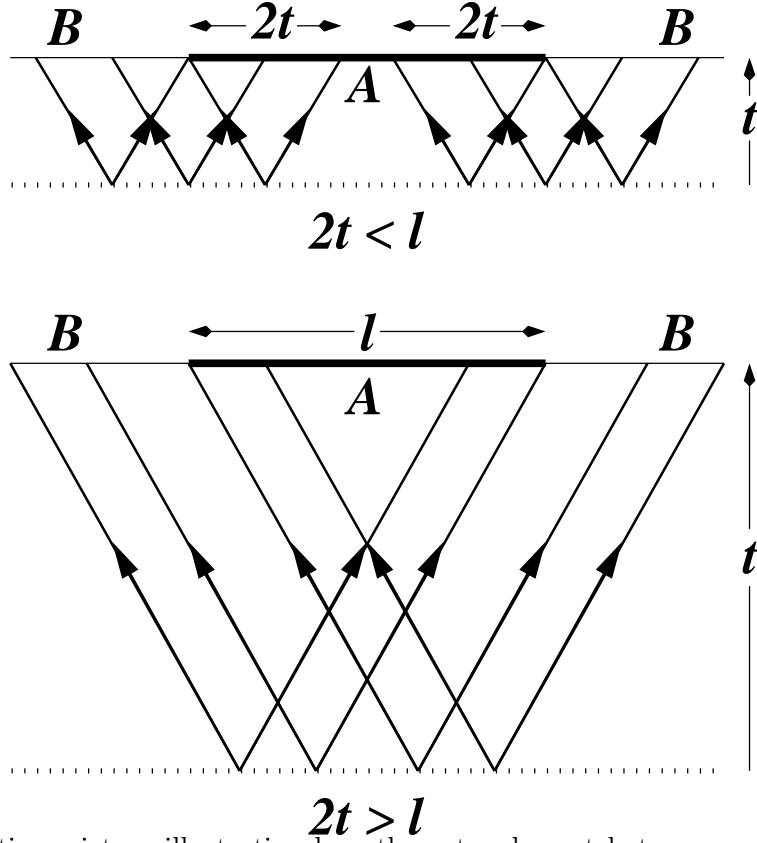}
\caption{\label{fig2tl}
Space-time picture illustrating how the entanglement between
an interval A and the rest of the system,
due to oppositely moving coherent quasiparticles, grows
linearly and then saturates. The case where the particles move only
along the light cones is shown here for clarity.}
\end{figure}

Now consider these quasiparticles as they reach either $A$ or $B$ at
time $t$. The field at some point $x'\in A$ will be entangled with that
at a point $x''\in B$ if a pair of entangled particles emitted from a
point $x$ arrive simultaneously at $x'$ and $x''$ (see Fig.~\ref{fig2tl}). 

The entanglement entropy between $x'$ and $x''$
is proportional to the length of the interval
in $x$ for which this can be satisfied. Thus the total entanglement
entropy is
\be
S_A(t)\approx \int_{x'\in A}dx'\int_{x''\in B}dx''\int_{-\infty}^\infty
dx\int
f(p',p'')dp'dp''\delta\big(x'-x-v(p')t\big)\delta\big(x''-x-v(p'')t\big).
\ee

Now specialize to the case where $A$ is an interval of length $\ell$.
The total entanglement is twice that
between $A$ and the real axis to the right of $A$, which corresponds to
taking $p'<0$, $p''>0$ in the above. The integrations over the
coordinates then give ${\rm max}\,\big((v(-p')+v(p''))t,\ell\big)$, so
that
\bea
S_A(t)&\approx& 2t\int_{-\infty}^0dp'\int_0^\infty dp''f(p',p'')
(v(-p')+v(p''))\,H(\ell-(v(-p')+v(p''))t)+
\nonumber\\
\nonumber\\&+&
2\ell \int_{-\infty}^0dp'\int_0^\infty dp''f(p',p'')
\,H((v(-p')+v(p''))t-\ell)\,,
\label{ppp}
\eea
where $H(x)=1$ if $x>0$ and zero otherwise. Now since $|v(p)|\leq 1$,
the second term cannot contribute if $t<t^*=\ell/2$, so that $S_A(t)$ is
strictly proportional to $t$. On the other hand as $t\to\infty$, the
first term is negligible (this assumes that $v(p)$ does not vanish
except at isolated points), and $S_A$ is asymptotically proportional to
$\ell$, as found earlier. 

However, unless $|v|=1$ everywhere (as is the case for the conformal
field theory calculation), $S_A$ is not strictly proportional
to $\ell$ for $t>t^*$. In fact, it is easy to see that the asymptotic
limit is always approached from below, as found for the Ising spin chain
in Sec.~\ref{secIs}. The rate of approach depends on the behavior of
$f(p',p'')$ in the regions where $v(-p')+v(p'')\to 0$. This generally
happens at the zone boundary, and, for a non-critical quench, also at 
$p'=p''=0$. If we assume that $f$ is non-zero in those regions, we find
a correction term $\sim -\ell^3/t^2$ in the limit where $t\gg t^*$. 

The above result may be generalized to the case when A consists of
several disjoint intervals, and checked against the corresponding
conformal invariance result in the case where $|v(p)|=1$. 
This is discussed in
Appendix \ref{slitapp}. In particular one should note that $S_A(t)$ is not
always non-decreasing. For example, if A consists of a regular
array of intervals, each of length $\ell$, separated by gaps also
of length $\ell$, the entanglement entropy oscillates in a
saw-tooth fashion with a period $\ell$.

\section{Discussion}
\label{disc}

We have presented results on the time-dependence of the entanglement
entropy $S_A$, 
starting from a translationally invariant pure state $|\psi_0\rangle$
which is
not an eigenstate of the time evolution operator, from two complementary
perspectives. The first was conformal field theory, which applies only
in the asymptotic regime, at the critical point, to conformally
invariant initial states, in theories with a
purely relativistic dispersion relation. In order to regularize these
calculations and maintain tractability we were forced to apply a
high energy cut-off $e^{-\e E}$ to the state. The second was a solvable
lattice model, the Ising model in a transverse field, in which we were
able to perform calculations starting from a variety of initial
states, considering both critical and non-critical quenches. In these
calculations the only dimensional microscopic parameter is the lattice
spacing. It should be noted that the continuum limit of the critical
transverse Ising spin chain is known to correspond to a conformal
field theory.

The results of these two sets of calculations have a number of important
features in common: the entanglement entropy increases linearly with
time $t$ (after transients die away in the lattice case), up to
$t^*=\ell/2$, in units where the maximum propagation speed of excitations
is taken to be unity. For $t\gg t^*$, $S_A/\ell$ saturates at an
asymptotic value. 

However, there are also a number of differences. Partly these may be
explained by
the set-up being different: in the CFT calculation the scale of
slope of the linear $t$-dependence, as well as the asymptotic value of
$S_A/\ell$, is set by $\e$, while for the lattice calculation it is set
by the lattice spacing as well as the parameter $h_0$ specifying the
initial state. However, in the CFT calculation the discontinuity in the
slope is rounded only over a time interval $O(\e)$, after which $S_A$
reaches its asymptotic value immediately. This means that the ratio
\begin{equation}
R\equiv\frac{(\partial S_A/\partial t)_{t<t^*}}{2
(\partial S_A/\partial\ell)_{t\gg t^*}}
\end{equation}
is exactly unity (in units where the maximum speed of propagation is 1.)
On the other hand, the lattice calculations exhibit a slow increase
towards the asymptotic value for $t>t^*$, so that $R$ is always less
than unity. In Sec.~\ref{arg} however, we argue that this can be
understood within a simple physical picture and the fact that, on the
lattice, there are quasiparticle excitations which travel with a group
velocity less than the maximum allowed value. This also accounts for the
existence of a sharp feature in $S_A(t)$ even for non-critical quenches
to $h>1$, which should correspond, in the scaling limit, to a 
relativistic theory with a mass gap.
In these cases we find the explicit formula 
\be
R= \frac{\int_{-\infty}^0 dp' \int_0^\infty dp'' f(p',p'')[v(-p')+v(p'')]}{
2\int_{-\infty}^0 dp' \int_0^\infty dp'' f(p',p'')}\leq 1\,.
\ee

There are however other puzzling features of the Ising results, which may
however be specific to this model, for example the facts that the
limiting behavior $\lim_{\ell\to\infty}S_A/\ell$ is a symmetric
function of $h_0$ and $h$, and that the detailed time-dependences of all the
quenches from $h_0=\infty$ to any value of $h>1$ all appear to collapse
all on the same curve when suitably rescaled. Clearly these 
deserve more investigation to see whether they persist in other models.

In this paper, we have restricted attention to quenches from a
translationally invariant initial state $|\psi_0\rangle$. It is also
interesting to consider other setups, for example when initially $A$ and
$B$ are completely decoupled and in their respective ground states. This
is currently under investigation. It should also be noted that the
causal picture of Sec.~(\ref{arg}) may be simply generalized to higher
dimensions, with results which are, however, more dependent on the
precise geometry.

Finally we should point out that the increase of the entanglement
entropy we have found
for a single interval $A$ in an infinite system is in no way
connected with the second law of thermodynamics. As shown in Appendix
\ref{slitapp}, in more complicated situations, it may decrease or even 
oscillate.

\section*{Acknowledgments}
We thank Rosario Fazio for bringing the problem of the dynamics of 
entanglement to our attention.
PC thanks Fabian Essler and Alex Lefevre for useful discussions and comments 
about Toeplitz determinants.
This work was supported in part by the EPSRC under Grant GR/R83712/01.
Part of it was carried out while JC was a member of the Institute
for Advanced Study.
He thanks the School of Mathematics and the School of Natural Sciences
for their hospitality.
This stay was supported by the Ellentuck Fund.

\appendix

\section{Computation of $S_\ell$ for the Ising chain}
\label{app}

The diagonalization of $H_I$ is presented in several standard textbooks 
(see e.g. \cite{sach}). Nevertheless we briefly present it here to make 
the notations self-explanatory.
The hamiltonian for a chain of length $N$ is 
\be
H_I = - \frac{1}{2}\sum_{l=-({N-1})/{2}}^{({N-1})/{2}} 
[ \sigma_l^x \sigma_{l+1}^x+ h\sigma_l^z]\,,
\label{HamI}
\ee
and we consider open boundary conditions. We are interested in the limit 
$N\rightarrow\infty$. 
The first step in diagonalizing the hamiltonian is to introduce a set of 
spinless fermion annihilation and creation operators through the 
non-local Jordan-Wigner transformations
\be
\Fa_l = \left( \prod_{m<l}\sigma_m^z \right) \frac{\sigma_l^x -
 i\sigma_l^y}{2}\,, 
\qquad
\Fa_l^{\dagger} = \left( \prod_{m<l}\sigma_m^z \right) \frac{\sigma_l^x +
 i\sigma_l^y}{2}\,,
\ee
where $\Fa_l^{\dagger}$ stands for the Hermitian conjugate of $\Fa_l$.
They satisfy the anticommutation relations
\be
\{\Fa_l^{\dagger},\Fa_m\}=\delta_{lm} , \qquad\{\Fa_l,\Fa_m\} =0.
\ee
Let us now introduce the Fourier transformed fermionic operators 
\be
\Fd_k =\frac{1}{\sqrt{N}} \sum_{l=-({N-1})/{2}}^{({N-1})/{2}} 
\Fa_l e^{-i\frac{2\pi}{N}kl},
\ee 
with $-(N-1)/2 \leq k \leq (N-1)/2$. Since this transformation
is unitary, the anticommutation relations  
$\{\Fd^{\dagger}_k,\Fd_p\}=\delta_{kp}$ remain valid.
The hamiltonian now takes the form 
\be
H_I=\sum_{k=-(N-1)/2}^{(N-1)/2} 
\left(\left[h-\cos \frac{2\pi k}{N}\right] \Fd^{\dagger}_k \Fd_k - 
\frac{i}{2} \sin\frac{2\pi k}{N} 
\left[\Fd_{-k}\Fd_k + \Fd^{\dagger}_{-k} \Fd^{\dagger}_k\right]-\frac{h}{2}
\right)\,.
\ee

A final unitary transformation is needed to cast the hamiltonian into
diagonal form. This is the so-called Bogoliubov transformation 
\be
\Fb^{\dagger}_k =u_k \, \Fd^{\dagger}_k + i v_k \, \Fd_{-k}\,, \qquad 
\Fb_k = u_k \, \Fd_k - i v_k \, \Fd^{\dagger}_{-k},
\label{bogo}
\ee 
where $u_k = \cos{\theta_k/2},~ v_k = \sin{\theta_k/2}$ and
\be 
\tan{\theta_k} = \frac{\sin{\frac{2\pi k}{N}}}{
\cos{\frac{2\pi}{N}k}-h}\,.
\label{thetak}
\ee
Again, due to unitarity of the Bogoliubov transformation, the operators
$\{\Fb_k\}$ follow the usual anticommutation relations 
$\{\Fb^{\dagger}_k,\Fb_p\}=\delta_{kp}$.  
Finally, the hamiltonian takes a diagonal form 
\be
H_I=\sum_{k=-(N-1)/2}^{(N-1)/2} \e_k \, (\Fb^{\dagger}_k \Fb_k-1/2),
\label{Hambog}
\ee
where
\be
\e_k \equiv \sqrt{\left(h-\cos{\frac{2\pi k}{N}}\right)^2+ 
\sin^2{\frac{2\pi k}{N}}}.
\label{ek}
\ee
The thermodynamic limit is obtained by defining $ \p =2\pi k/N$
and sending $N\to \infty$, giving
\be 
H_I = \int^{\pi}_{-\pi}\frac{{\rm d}\p}{2\pi}\ \e_\p \Fb^{\dagger}_\p \Fb_\p, 
\ee 
with 
\be
\label{energy}
\e^2_\p = (h-\cos{\p})^2 +  \sin^2{\p}\,,  
\ee
where we omit a constant term from the hamiltonian.

For the time evolution, we closely follow Ref. \cite{sps-04}, whose 
results are mainly based on some previous works \cite{bm-72,ir-00}. 
Since the hamiltonian is translational invariant,
only fermionic states with opposite momentum $k$ and $-k$
are mixed. We may therefore write the two component column vectors
\be
\FB_k =
\begin{pmatrix}
\Fb_k\\
\Fb_{-k}^\dag
\end{pmatrix}\,,
\qquad\rm{and} \qquad
\FD_k =
\begin{pmatrix}
\Fd_k\\
\Fd_{-k}^\dag
\end{pmatrix}\,.
\ee 
The Bogoliubov transformation relating $\FB_k$ and $\FD_k$ is
expressed as $\FD_k = R_x(\theta_k) \FB_k$, where 
\be 
R_\mu(\alpha)= \cos\frac{\alpha}{2} + i\sigma_\mu\sin\frac{\alpha}{2} 
\ee 
and $\sigma_\mu$ are the Pauli matrices with $\mu=x,y,z$. (These are used for
conciseness of notation and should not be confused with the
operators representing the `spins' of the Ising chain.)

For $t < 0$, the system is taken to be in the ground state corresponding to 
the field $h_0$.
It is natural to define $\Fb'$ fermions as those which
diagonalize the hamiltonian with $h_0$. 
Similarly, $\theta_k'$ and $\FB_k'$ are given by
analogy with (\ref{thetak}) and (\ref{bogo}) with $h$ replaced by $h_0$.
The state $|\psi_0\rangle$ is therefore the vacuum of $\FB'$
fermions. This in particular implies 
\be 
\langle\psi_0 |
B_k' B_k'^\dag|\psi_0\rangle =
\begin{pmatrix}
1 & 0\\
0 & 0
\end{pmatrix}
= \frac{1}{2}(\sigma^z + 1)\,,
\ee 
that in terms of $B_k$ fermions reads
\be 
\langle\psi_0| B_k B_k^\dag|\psi_0\rangle = 
R_{x}(\theta'_k -\theta_k) \frac{1}{2}(\sigma^z + 1) 
R_x^\dag(\theta'_k - \theta_k)=
\frac{1}{2}(1 + \sigma^z \cos\phi_k + \sigma^y \sin \phi_k), 
\ee 
with $\phi_k = \theta'_k - \theta_k$.

The time evolution of the operators now proceeds according to the hamiltonian
(\ref{Hambog}), so that $B_k(t) = U_k(t) B_k(0)$, where 
\be 
U_k(t) \equiv e^{-i H_I t}=
\begin{pmatrix}
e^{-i \e_k t} & 0\\
0 & e^{i \e_k t}
\end{pmatrix}
= R_z(-2\e_k t). 
\ee 
The expectation values at any time can be easily evaluated.
Let us introduce the operators 
\be 
\o^+_i=\Fa^\dag_i+\Fa_i\,,\qquad {\rm and} \qquad 
\o^-_i=\Fa^\dag_i-\Fa_i\,,
\ee
and the two-component vector operator
\be 
\Omega_k(t) = \begin{pmatrix}
\o^+_k(t)\\\o^-_k(t)
\end{pmatrix}
= \sqrt{2} R_{y}\left(\frac{\pi}{2}\right) \FD_k(t), 
\ee 
where we defined $\o^{\pm}_k$ as the Fourier transform of 
$\o^{\pm}_i$. 
Using the algebra of $SU(2)$ matrices, it is straightforward to 
find \cite{sps-04}
\bea 
\langle\psi_0 | \Omega_k \Omega_k^\dag|\psi_0\rangle &=&
\langle\psi_0 |
\begin{pmatrix}
\o^+_k \o^+_{-k} & -\o^+_k \o^-_{-k} \nonumber\\
\o^-_k \o^+_{-k} & -\o^-_k \o^-_{-k}
\end{pmatrix}
|\psi_0\rangle =\\
&=& \begin{pmatrix}
1 - \sin\phi_k\sin 2\e_k t & -e^{i \theta_k} (\cos \phi_k + i \sin\phi_k\cos 2\e_k t)\\
-e^{-i \theta_k} (\cos \phi_k - i \sin\phi_k \cos 2\e_k t) &
1 + \sin\phi_k\sin 2\e_k t
\end{pmatrix}.
\eea
Transforming back to real space we have 
\bea
\langle \o^+_l\o^+_j\rangle &=& 
\frac{1}{N}\sum_k e^{ik(l-j)}(1-\sin\phi_k\sin 2\e_k t)\,,\nonumber\\
\langle \o^-_l\o^-_j\rangle &=& 
\frac{1}{N}\sum_k e^{ik(l-j)}(-1-\sin\phi_k\sin 2\e_k t)\,,\nonumber\\
\langle \o^-_l\o^+_j\rangle &=& 
\frac{1}{N}\sum_k e^{ik(l-j)}e^{-i\theta_k}(-\cos\phi_k+ i\sin\phi_k\cos 2\e_k t)\,. 
\label{timeev}
\eea
Taking the thermodynamic limit, the sum over $k$ becomes 
an integral over $\p$ and
\bea
\tan\theta_k&\rightarrow& \tan\theta_\p= \frac{\sin\p}{\cos\p-h}\,,\nonumber\\
\phi_k&\rightarrow& \Phi_\p= \theta'_\p-\theta_\p\,, 
\eea
and $\e_k\rightarrow\e_\p$ defined by Equation (\ref{energy}).

As stated in the text, it is useful to introduce two Majorana 
operators \cite{Vidal} at each site of the spin chain 
according to equation (\ref{eq:aa}).
Operators $\Ma_m$ are Hermitian and obey anti-commutation relations, 
\be
\Ma_m^{\dagger} = \Ma_m,~~~~~~\{\Ma_m, \Ma_n\}= 2\delta_{mn}.
\ee
In terms of the Dirac operators introduced by the Jordan-Wigner 
transformation they read
\bea
\Ma_{2l-1}=\Fa^\dag_l+\Fa_l,&\qquad& \Ma_{2l}=-i(\Fa^\dag_l-\Fa_l)
\,,\nonumber\\
\Fa_l=\frac{\Ma_{2l-1}-i\Ma_{2l}}{2},&\qquad&
\Fa_l^\dag=\frac{\Ma_{2l-1}+i\Ma_{2l}}{2}\,,
\eea
and so, in terms of $\o^\pm$ they are
\be
\Ma_{2l-1}=\o^+_l\,,\qquad \Ma_{2l}=-i\o^-_l\,.
\ee
The two point time-dependent expectation values of $\Ma$ are easily 
obtained in terms of Eqs. (\ref{timeev}) as
\bea
\langle \Ma_{2l} \Ma_{2m}\rangle&=&-\langle\o^-_l\o^-_m\rangle\,,\nonumber\\
\langle \Ma_{2l-1} \Ma_{2m-1}\rangle&=&\langle\o^+_l\o^+_m\rangle\,,\nonumber\\
\langle \Ma_{2l} \Ma_{2m-1}\rangle&=&-i\langle\o^-_l\o^+_m\rangle\,,\nonumber\\
\langle \Ma_{2l-1} \Ma_{2m}\rangle&=&-i\langle\o^+_l\o^-_m\rangle\,.
\label{majoex}
\eea

The entanglement entropy for the time-dependent state is now obtained 
exactly as in the ground-state, following closely Ref. \cite{Vidal}.
From equation ({\ref{majoex}), the time dependent 
correlation matrix $\ev{\Ma_m\Ma_n} = \delta_{mn} + i \Gamma^A_{mn}$ 
of the Majorana operators $\Ma$ reads
\be
\Gamma^A = \left[
 \begin{array}{ccccc}
\Pi_0  & \Pi_{-1}   &   \cdots & \Pi_{1-N}  \\
\Pi_1 & \Pi_0   & &\vdots\\

\vdots&  & \ddots&\vdots  \\
\Pi_{N-1}& \cdots  & \cdots  & \Pi_0 
\end{array}
\right], ~~~ \Pi_l = \left[\begin{array}{cc}
-f_l & g_l \\
-g_{-l} &  f_l
\end{array}
\right]\,.
\label{eq:GammaA}
\ee
The coefficients $g_l$ and $f_l$, in the limit of an infinite 
chain are easily derived from equations 
(\ref{timeev}) and (\ref{majoex}), obtaining the results reported in the text
equation (\ref{eq:g2}).

As shown in Ref. \cite{Vidal},
the entropy of the reduced density matrix $\rho$ for $\ell$ adjacent spins
can be computed from $\Gamma^A$, in spite of the non-local character of 
transformation (\ref{eq:aa}).
Indeed, the symmetry  
\be
\left(\prod_l \sigma_l^z \right) H_I \left(\prod_l \sigma_l^z \right) = H_I
\ee
implies that the trace over a given spin at the site $j$ is equivalent to 
the trace over the two Majorana operators at the same site.  
Thus the density matrix $\rho_\ell$ can be reconstructed from the restricted 
$2\ell\times 2\ell$ correlation matrix
\be
\ev{\Ma_m\Ma_n} = \delta_{mn} + i(\Gamma^A_\ell)_{mn},~~~~~m,n =1,\cdots,2\ell,
\ee
where $\Gamma^A_\ell$ is the matrix obtained by $\Gamma^A$ cutting the 
last $N-\ell$ rows and columns.

An orthogonal matrix $V$ bring $\Gamma^A_\ell$ into the block-diagonal form
\be
V\Gamma^A_\ell V^T = \bigoplus_{l=1}^\ell \nu_l
\left[ \begin{array}{cc} 0 &1  \\
-1 &0 \end{array}
\right].
\label{eq:GammaC}
\ee
This means that the reduced density matrix can be written as a product
$\rho_\ell=\varrho_1 \otimes \cdots \otimes \varrho_\ell$,
where each $\varrho_l$ has eigenvalues $(1\pm\nu_l)/2$ and entropy
$S(\varrho_l)= H(\nu_l)$, where $H(x)$ is given by equation (\ref{Hx}).
Thus the entropy of $\rho_\ell$ is the sum of entropies of the $\ell$ 
uncorrelated modes,
\be
S_\ell = \sum_{l=1}^\ell H(\nu_l).
\label{eq:entronu}
\ee

Summarizing: for arbitrary time, and in the thermodynamic limit 
($N\rightarrow \infty$), the time evolution of the entropy $S_\ell$ of the
Ising model can be obtained by 
\begin{itemize}
\item evaluating Eq. (\ref{eq:g2}) numerically for $l=0,\cdots, \ell-1$, 
\item diagonalizing $\Gamma_\ell^A$ to obtain $\nu_l$, and 
\item evaluating $S_\ell$ using Eq. (\ref{eq:entronu}).
\end{itemize}

\section{Some analytic results for the correlation matrix}
\label{appb}

In some cases it is possible to get analytic expression for the coefficients 
$g_l$ and $f_l$, that help the numerical computation in the general case.
For example, in the limit $h_0\rightarrow\infty$ and for $t\rightarrow\infty$
we get
\be
g_l=\frac{1}{2\pi} \int_0^{2\pi} d\p e^{-i\p l} 
\frac{(e^{-i\p}-h)}{\e_\p^2} (h-\cos\p)=
\begin{cases}
\frac{1-h^2}{2h^{l+2}}\qquad l\geq1\\
\frac{1}{2h^2}-1\qquad l=0\\
\frac{1}{2h}\qquad l=-1\\
0\qquad l\leq-2
\end{cases}\,.
\label{glh0infinity}
\ee
Using these values in equation (\ref{eq:g2}) reduces the difficulty of the 
numerical calculations for generic times.
 
For the a critical quench ($h_0=\infty$ and $h=1$), the infinite time
value of $g_l$ reduces to
\be
g_l=\frac{1}{4\pi}\int_0^{2\pi} d\p [e^{-i\p (l+1)}-e^{-i\p l}]=
\frac{\delta_{l,-1}-\delta_{l,0}}{2}
\ee
In this case, the Toeplitz matrix has only the two diagonals closest to the 
principal with non-zero entries $\pm1/2$.
This matrix is nothing but the Laplacian on the lattice (minus 1, and without
two elements at $(2\ell,1)$ and $(1,2\ell)$ that for $\ell\rightarrow\infty$ are 
irrelevant). Thus the eigenvalues for large $\ell$ are 
\be
\nu_m=\cos\frac{\pi m}{2\ell}\qquad m=1,\dots \ell\,,
\ee
and the entropy is
\be
\lim_{\ell\rightarrow\infty}\frac{S_\ell(\infty)}{\ell}=
\frac{1}{\ell}\sum_{m=1}^\ell H(\cos\frac{\pi m}{2\ell})\rightarrow
\frac{2}{\pi}\int_0^\pi d y H(\cos y)=
 2\left(\log2-\frac{1}{2}\right) \,.
\ee

\section{General result for an arbitrary number of intervals.} 
\label{slitapp}

In this appendix we sketch the CFT calculation of
$S_A(t)$ for the case when $A$ consists of the union of the $N$
intervals $(u_{2j-1},u_{2j})$ where $1\leq j\leq N$ and
$u_k<u_{k+1}$. ${\rm Tr}\,\rho_A^n$ is given as usual by the ratio
$Z_n/Z_1^n$ which has the form of a correlation function
\begin{equation}
\label{cf12}
\langle\prod_j\Phi_n(u_{2j-1}+i\tau_1)
\prod_j\Phi_{-n}(u_{2j}+i\tau_1)\rangle\,,
\end{equation}
in a strip of width $2\epsilon$. This is related to the upper half
$z$-plane by $w=(2\epsilon/\pi)\log z$, whereby the points
$u_k+i\tau_1$ are mapped to $z_k=e^{\pi(u_k+i\tau_1)/2\epsilon}$.
If we choose $\sum_ku_k=0$, the jacobian factors in the
transformation are constant, independent of the $u_k$.
Analytically continuing to $\tau_1=\epsilon+it$ sets
$z_k=ie^{\pi(u_k-t)/2\epsilon}$, that is, on the imaginary axis,
while the image points go over into $\bar
z_k=-ie^{\pi(u_k+t)/2\epsilon}$. The correlation function
(\ref{cf12}) is a generalization of that in (\ref{zzzz}) and is a
product of several factors: the first involves differences
$(z_k-z_l)$ raised to the power $\pm 2n\Delta_n$ according to
whether $k-l$ is even or odd; the second is the same with $z_k$
replaced by $\bar z_k$; and the third involves factors like
$(z_k-\bar z_l)$. It can be seen that the $t$-dependence cancels
between the first and second type of terms. On the other hand
\begin{equation}
|z_k-\bar z_l|=e^{\pi(u_k-t)/2\epsilon}+ e^{\pi(u_l+t)/2\epsilon}
\sim e^{\pi{\rm max}(u_k-t,u_l+t)/2\epsilon}\,,
\end{equation}
where we have assumed that $u_k\pm t\gg\epsilon$. This expression
is to be raised to the power $\pm 2\Delta_n$ according to whether
$k-l$ is odd or even. Finally differentiating with respect to $n$
and setting $n=1$ we find
\begin{equation}
S_A(t)\sim S_A(\infty) +\frac{\pi c}{12\e}
\sum_{k,l}(-1)^{k-l-1}{\rm max}(u_k-t,u_l+t)\,.
\end{equation}
If $N$ is finite (or more generally the $u_k$ are bounded) the
second term vanishes for sufficiently large $t$. At shorter times,
$S_A(t)$ exhibits piecewise linear behavior in $t$ with cusps
whenever $2t=u_k-u_l$, at which the slope changes by $\pm(\pi
c/6\epsilon$ according to whether $k-l$ is even or odd. In the
case of an infinite number of regular intervals, with $u_k=k\ell$,
$k\in{\rm Z}$, $S_A(t)$ exhibits a sawtooth behavior.

\end{document}